\documentclass[superscriptaddress,twocolumn,prl]{revtex4}
\usepackage{graphicx}
\usepackage{dcolumn}
\usepackage{bm}
\usepackage{color}
\usepackage{ulem}
\usepackage{amsmath}
\usepackage{amsfonts}
\usepackage{amssymb}
\usepackage{tikz}
\usepackage[final]{changes}

\definechangesauthor[name={2nd}, color=orange]{2nd}
\begin{document}

\title{Stability of even-denominator fractional quantum Hall states in systems with 
strong Landau-level mixing}
\author{Wenchen Luo}
\affiliation{School of Physics and Electronics, Central South University, Changsha, China 410083}
\author{Shenglin Peng}
\affiliation{School of Information Science and Technology, Northwest University, Xi'an,
China 710127}
\affiliation{School of Physics and Electronics, Central South University, Changsha,
China 410083}
\author{Hao Wang}
\affiliation{Shenzhen Institute for Quantum Science and Engineering, 
Southern University of Science and Technology, Shenzhen, China 518055}
\author{Yu Zhou}
\email{yzhou@just.edu.cn}
\affiliation{Department of Physics, Jiangsu University of Science and Technology,
Jiangsu, China 212003}
\author{Tapash Chakraborty }
\email{Tapash.Chakraborty@Umanitoba.ca}
\affiliation{Department of Physics, Brock University, St. Catharines, ON, 
Canada L2S 3A1}
\affiliation{Department of Physics and Astronomy, University of Manitoba, Winnipeg, Canada R3T 2N2}
\date{\today }

\begin{abstract}

Mixing of Landau levels has been understood to be essential in governing the nature of the 
ground state for the even-denominator fractional quantum Hall effect. The incompressibility 
of the ground state at filling factor $5/2$ in the strong Landau-level-mixed systems, such 
as the ZnO quantum well, is not always stable. Here we present an approach to generally deal 
with this kind of systems and satisfactorily explain the recent experiments 
[Falson \textit{et al}. Sci. Adv. \textbf{4}, eaat8742 (2018)] by implementing 
the screening plus the thickness effect. Further, the phase diagrams of the incompressibility 
of the ground state indicate that the phase transitions can be explicitly extracted by observing 
the lowest gap of the collective modes when the magnetic field and the width of the quantum well
are tuned. We also predict the incompressibility of the two-dimensional electron gas in higher 
Landau levels in another strong Landau-level-mixed system, viz., the black phosphorene, by 
considering the screening effect where the relevant even-denominator fractional quantum Hall 
effects can possibly be observed.
\end{abstract}

\maketitle


For the past four decades,  the quantum Hall effects have been 
the epitome of elegant phenomena in condensed matter physics \cite{review}. 
Ever since the observation of the even-denominator fractional quantum Hall effect (FQHE) 
\cite{odfqhe}, it has been recognized that some difficulties remain in explaining this effect.
Experimentally, the spin polarization of the 
$5/2$ FQHE state is still debated \cite{argue} and very high-quality samples are an absolute 
requirement to resolve the issue. A couple of trial wave functions have been proposed thus far for 
this system \cite{mr,trialwf}. However, the numerical results are somewhat size
dependent so that the trial wave function candidates are not so satisfactory and show relatively 
poor performance when compared with the numerical wave functions. It is also possible that both 
the long-range and the short-range interactions are important but the Haldane's pseudopotentials 
are limited to a spherical geometry. In a toroidal geometry the numerical strategy, although 
not entirely trouble free, can be a powerful alternative approach.

In practice, the even-denominator FQHE is proposed to be useful in topological quantum
computation \cite{nonab} since the non-Abelian excitations in the even-denominator FQHE are 
robust against many environment noises due to the topological protection. Determining the nature 
of the even-denominator FQHE states is therefore important for development of quantum computation. 
The Moore-Read pfaffian trial wave function (or its particle-hole conjugate anti-pfaffian state) is
the most plausible candidate for the ground state, albeit the overlap between this wave function 
and the numerical results is not being high. A particle-hole symmetric pfaffian state for the 
ground state is proposed \cite{son} that is believed to be stable when the Landau level (LL) mixing 
and the influence of disorders in the two-dimensional electron gas (2DEG) are included 
\cite{feldman}. Further studies have indicated the existence of a phase transition when the LL 
mixing increases \cite{luo3}. The incompressibility of the 2DEG depends on the strength of the LL 
mixing which can be effectively considered by screening the Coulomb potential. In the framework of 
screening, the surprise missing of the 5/2 plateau in ZnO quantum well has been explained 
\cite{falson,luo1}. In a more recent experiment performed in 2018, however, Falson \textit{et al} 
reported that the $5/2$ FQHE is recovered in one sample \cite{falson2}, which could not be explained 
by our previous theoretical approach. In order to understand the strong LL mixed 2DEG in 
general, and why the $5/2$ FQHE is survived in one sample but is still missing in another sample 
of the 2018 experiment, we need to consider the screening effect together with the thickness effect 
associated with the quantum well in our analysis below.

The 2DEG discussed here has a unique property that the LL gap is very small compared to the 
Coulomb energy gap. The LL mixing is therefore strong and influences the properties of the ground 
state, and its incompressibility may be absent. A dimensionless quantity is defined 
to characterize the LL mixing strength: the ratio of the Coulomb interaction strength 
$E^{}_{\rm C}=e^2/{\epsilon\ell}$ to the LL gap $E^{}_{\rm cyc} =\hbar \omega^{}_c$, $\kappa
=E^{}_{\rm C}/E^{}_{\rm cyc}$, where $\epsilon$ is the dielectric constant of the material,
$\ell =25.656/\sqrt{B}$ nm is the magnetic length, and $\omega^{}_c$ is the cyclotron frequency 
in the magnetic field $B$. This ratio is typically small in a conventional GaAs quantum well 
in a strong magnetic field, and is a constant $0.5\sim 0.8$ in graphene \cite{graphene}. In 
contrast, it is very large in ZnO, $\kappa >10$ (the same order as in black phosphorene). The 
perturbation theory approach, including the 3-body interaction which is based on the expansion 
of $\kappa $ \cite{perturbation} is not appropriate for these systems. To overcome 
this difficulty, we make use of the screened Coulomb potential in the relevant LL, which is 
obtained by integrating out all the other LLs, to effectively describe the LL mixing.


As has been done in previous works \cite{luo1}, we also consider the electron-electron 
interaction described by the screened Coulomb potential in our present approach. However, if we 
just consider the pure 2DEG, we only have unstable $5/2$ FQHE in both systems considered
in the experiment \cite{falson2,joe_more}, which clearly conflicts with the experimental observations.
It has been suggested that we need to consider the third dimension, the thickness, of the quantum 
well in order to get a better understanding of the present situation. For simplicity, the 
confinement of the quantum well is approximately described either by a parabolic potential
\cite{parabolic,luo2} or an infinite square potential \cite{peterson}. We will compare these 
two approximations, of which the results are supposed to be coincident.

The many-body Hamiltonian is given by
\begin{eqnarray}
&&H=\sum_{n,i,\sigma}E^{}_{n,\sigma }c_{n,i,\sigma }^{\dag }c^{}_{n,i,\sigma}+
\frac12\sum_{\sigma,\sigma^{\prime}}\sum_{\substack{n^{}_1,\ldots,n^{}_4  \\ 
i^{}_1,\ldots,i^{}_4}} V_{\left(s\right),i^{}_1,i^{}_2,i^{}_3,i^{}_4}^{n^{}_1,n^{}_2,
n^{}_3,n^{}_4}  \notag \\
&&\times\, c_{n^{}_1,i^{}_1,\sigma}^{\dag}c_{n^{}_2,i^{}_2,\sigma^{\prime
}}^{\dag }c^{}_{n^{}_3,i^{}_3,\sigma^{\prime}}c^{}_{n^{}_4,i^{}_4,\sigma},
\end{eqnarray}%
where $c$ is the electron operator, $E^{}_{n,\sigma}$ is the kinetic energy of
the LL $n$ with spin $\sigma $, $n^{}_i$ is the LL index and $i^{}_{1,\ldots,4}$
are the guiding center indices. The Coulomb interaction matrix element 
$V_{\left(s\right),i^{}_1,i^{}_2,i^{}_3,i^{}_4}^{n^{}_1,n^{}_2,n^{}_3,n^{}_4}$
depends on the confinement potential, which will be explicitly given in the
following cases. We perform the exact diagonalization scheme to numerically
solve the Hamiltonian with the translational symmetry in the toroidal
geometry \cite{haldane,book,exactdiag}. The collective modes of the system at
zero temperature are obtained, and the incompressibility or instability of the
FQHE state is then determined.

The virtual process between these LLs in the density response function should be excluded to 
avoid double counting of the correlations if more LLs are involved in the Hamiltonian. In the 
following, we need to analyze the screened dielectric functions in detail for different 
confinement potentials.


The non-interacting Hamiltonian can be exactly solved in a parabolic
potential in a tilted (or perpendicular) magnetic field,
and the screened Coulomb potential has been studied \cite{luo2}. We will
numerically evaluate the collective modes in different magnetic fields and
different widths of the quantum well based on the formulations of Ref.~\cite{luo2}.

For an infinite square potential, the non-interacting
Hamiltonian can also be exactly solved when the magnetic field is
perpendicular to the $xy$ plane, in which the $z$ component of the wave
function can be separated, $\Psi^{}_{m,n,X}\left(\mathbf{r},z\right)
=\sqrt{2/ L^{}_z}\sin \left( m \pi z / L^{}_z \right) \psi _{n,X}\left(
\mathbf{r}\right)$, where $L^{}_z$ is the width of the quantum well, $m$ is
the Landau band index, $n$ is the LL\ index, $X$ is the guiding center index,
and $\psi_{n,X}\left(\mathbf{r}\right)$ is the wave function of a
conventional 2DEG in a magnetic field in the Landau gauge.

The screened dielectric function is given by
\begin{equation}
\epsilon^{}_s\left(q\mathbf{,}q^{}_z\right) =1-\frac{4\pi e^2}{\left(
q^2+q_z^2\right)\epsilon}\chi_{nn}^0\left(q\mathbf{,} q^{}_z\right),
\end{equation}%
where the three-dimensional momentum contains the in-plane momentum $q$ and the 
$z$-component momentum $q^{}_z$. $\chi_{nn}^0\left(q\mathbf{,}q^{}_z\right)$
is the noninteracting retarded density-density response function computed in
the random phase approximation in the static limit \cite{sm}. If we consider that the
relevant LLs are all in the lowest band, i.e., $m^{}_i=1$, then the
screened Coulomb interaction matrix element is
\begin{eqnarray}
&&V_{\left(s\right), i^{}_1,i^{}_2,i^{}_3,i^{}_4}^{n^{}_1,n^{}_2,n^{}_3,n^{}_4}=
\frac{e^2}{\epsilon \ell}\frac{2}{\pi N^{}_s}\overline{\sum_{\mathbf{q}}}
V_{i^{}_1,i^{}_2,i^{}_3,i^{}_4}^{n^{}_1,n^{}_2,n^{}_3,n^{}_4}\left( \mathbf{q}\right)
\\
&&\times\, \int_0^{\infty }\frac{dq^{}_z\ell}{\epsilon^{}_s\left(
q,q^{}_z\right) \left( q^2+q_z^2\right) \ell^2}\frac{32\pi^4\left[
1-\cos \left( q^{}_zL^{}_z\right) \right]}{\left( 4\pi^2q^{}_zL^{}_z-q_z^3L_z^3\right)^2},  \notag
\end{eqnarray}
where $\overline{\sum}$ means that the term of $q=0$ is excluded in the
summation, $V_{i_1,i_2,i_3,i_4}^{n_1,n_2,n_3,n_4}\left( \mathbf{q}
\right)$ 
is the Coulomb interaction matrix element for a 2DEG in a conventional
zero-width quantum well without screening \cite{sm}. The integral includes the
width and screening corrections.



\begin{figure}[ptb]
\centering
\includegraphics[scale=0.67]{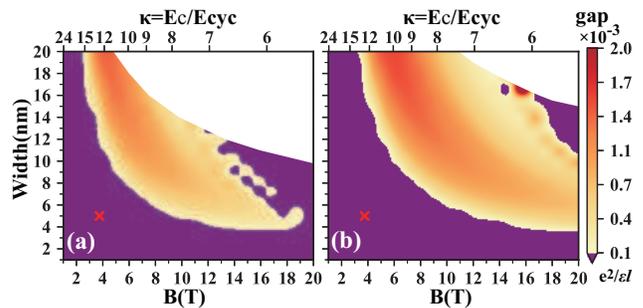}
\caption{The lowest energy gaps in the collective modes for different magnetic fields and 
widths of the wave functions in the ZnO quantum wells. Other parameters are extracted from Ref.~\cite{falson}, $m^*=0.44 m^{}_e$ 
and $\epsilon=8.5$, and $N_e=11$.  The confinement of the 2DEG in the $z$ direction is (a) parabolic potential 
and (b) infinite square well. When the gap is below $10^{-4}\, {e^2}/{\epsilon\ell}$ (could be negative), 
which is shown in the purple region, the 2DEG is not incompressible. The crosses mark the sample in 
Ref.~\cite{falson}, where $5/2$ FQHE is missing. }
\label{fig1}
\end{figure}

Let us check the theory to see whether it agrees with the experiment \cite{falson} by
using the parameters: half width of the wave function ($W=L_z/1.5$) about $5 - 6$ nm, dielectric constant
$\epsilon=8.5$, effective mass of the electron $m^* = 0.44 m^{}_e$ with 
the electron mass $m^{}_e$ . We first assume the incompressible phase as the ground state,
and evaluate the lowest excitation gap in the collective modes separating the
incompressible ground state from the excitations. If the gap is larger than $10^{-4}\,
e^2/{\epsilon \ell}$, then the FQHE is supposed to be stable. The reason for this particular
choice of $10^{-4}$ is because this energy corresponds to about $20$ mK, which  guarantees that the thermal fluctuation can not overwhelm the
incompressibility of the system, as otherwise the FQHE can not be observed in the
current experimental condition. It is clearly shown in Fig.~\ref{fig1} that the collective modes (at the
crosses) are softened at $B=3.75$ T and the translational invariant liquid phase no longer has the lowest 
energy. The 2DEG is compressible no matter what kind of potential is
chosen, which coincides with the experimental observation and the previous theoretic work \cite{luo1}.
Figure~\ref{fig1} also shows the phase diagram of the ground state at $\nu=5/2$ for
different values of magnetic fields and widths of the wave function, since the qualitative variation
of the gap must represent the changes in the ground states. The phase
diagrams for the two different potentials are qualitatively similar, i.e., the 5/2 FQHE is
only stable when the quantum well is thick and the LL mixing is weak (small 
$\kappa$). Note that the LLs are crossing and the noninteracting ground states are 
changed in the white regions in Fig. \ref{fig1}.

\begin{figure}[ptb]
\centering
\includegraphics[scale=0.65]{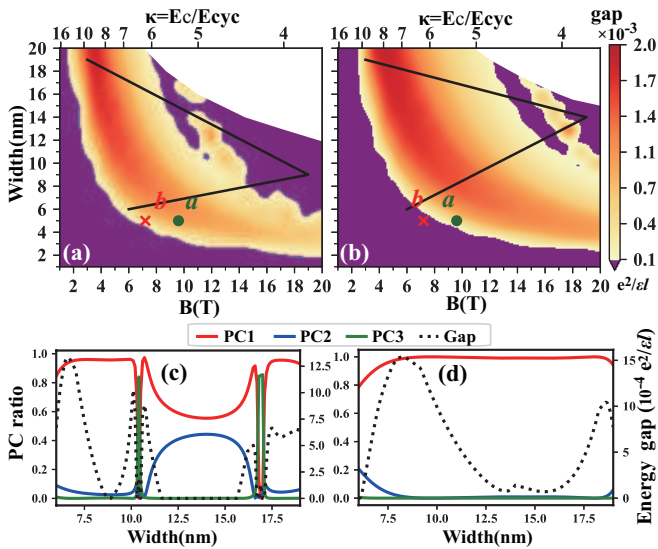}
\caption{Lowest gaps in the collective modes for different magnetic fields and the
widths of the wave functions in the ZnO quantum well with $m^*=0.3 m^{}_e$ and $\epsilon=8.5$ represent
the phase diagram in the strong LL mixed system. The confinement of the 2DEG in the $z$ direction 
is (a) a parabolic potential and (b) an infinite square well. The left cross and the right dot mark the 
phase of samples b and a in Ref.~\cite{falson2}, respectively. The first three principle components of the wave functions, which are extracted from the lines shown in (a) and (b), respectively, are shown in (c) and (d).}
\label{fig2}
\end{figure}

Let us now move on to the new experiments \cite{falson2} where the effective
mass is $m^*=0.3 m^{}_e$ at magnetic fields $B= 9.6$ T (sample a) and $7.2$ T
(sample b), corresponding to $\kappa = 5.4$ and $6.2$, respectively. 
The perturbation theory is still not very useful. Again, we compute the phase
diagrams for this effective mass in parabolic and infinite square potentials, shown in Fig. \ref{fig2}.
It is clear that for the infinite square well, we can quantitatively find that the 5/2
FQHE is stable in sample a and unstable in sample b (in Ref.~\cite{falson2}), while the 
incompressibility of the 2DEG can be obtained qualitatively in the parabolic potential. 
It implies that the infinite square well may be a better approximation to describe the quantum well.
Moreover, the second incompressible states found in tilted magnetic field in Ref.~\cite{falson2} 
can be explained by the phase transition induced by the LLs crossing when
the Zeeman coupling is in excess of the LL gap \cite{luo2}.

The phase diagrams are not qualitatively different from those of $m^*=0.44 m^{}_e$.
Some important features can be found here. First, there is an arc-like region isolated around the compressible region 
where the gaps become smaller or even negative for both types of potentials. 
It is similar to the case discussed in Ref.~\cite{luo3,dopedgaas} where a
topological phase transition was suggested. Here we also observe that the LL
mixing strength causes the phase transition. Besides, the width of the quantum
well is even involved to induce more than one phase transitions, which is not
expected earlier. Roughly, when $B \cdot W < 30$ T$\cdot$nm where $W$ is the 
wave function width, the 2DEG is compressible.
Second, the energy gaps are somewhat size-dependent. The gap becomes smaller when 
the electron number is increased. However, dealing with a larger number of electrons is not currently 
feasible, and we expect that for larger systems the FQHE in sample a still survives. We examine its
stability in a 3-LL model, i.e., we consider the Hamiltonian with three LLs
$N=0,1,2$ with $N^{}_e=15$ and the virtual process in these three LLs are
excluded in the screening. The collective mode clearly supports that the
ground state is incompressible.

To understand more about the incompressible phase, we perform the principle component 
(PC) analyze \cite{pca} to the ground state at zero momentum, as shown in Figs. \ref{fig2}(c) 
and (d). The wave functions are extracted along the black lines in Figs. \ref{fig2}(a) and (b). 
The first PC is almost the same in the arc region, which means that the ground states 
in the arc region share the same phase.


We next report on our study of black phosphorene, another strong LL mixed system. Here we need
to consider its bilayer structure in which the inter-layer Coulomb potential is different from 
the intra-layer Coulomb potential. We are required to extend the formula of screening to a bilayer
system, i.e., the Coulomb potential and the density response function should be
replaced by matrices,
\begin{equation}
V^s(q)=\left(
\begin{array}{cc}
V^s_{11}(q) &V^s_{12}(q) \\
V^s_{21}(q) & V^s_{22}(q)
\end{array}
\right),
\end{equation}
where $V^s_{kl}(q)$ is the screened Coulomb potential between layers $k$ and $l$ 
\cite{sm}.

Black phosphorene has a rectangular crystal lattice. There are four
atoms $A,B,C,D$ in a unit cell in which $A,B$ are in layer 1 and $C,D$ are
in layer 2. It can be simplified to a two-band model when we only work on
the bands near the Fermi surface \cite{bptwobands}. The positive and negative
filling factors correspond to the conduction and the valance bands, respectively.

The single-particle wave function is written as
\begin{equation}
\psi_{n,X}^{bp}=\frac1{\sqrt2}\sum_{m}\left(
\begin{array}{c}
u^{}_{n,m}\psi^{}_{m,X}\left( \left\vert B\right\rangle +\left\vert
C\right\rangle \right) \\
v^{}_{n,m}\psi^{}_{m,X}\left( \left\vert A\right\rangle +\left\vert
D\right\rangle \right)
\end{array}
\right),
\end{equation}
where $u^{}_{n,m}$ and $v^{}_{n,m}$ can be obtained by diagonalizing the
non-interacting Hamiltonian which is given in Refs.~\cite{bptwobands,bpfqhe}.
The Coulomb interaction matrix element reads
\begin{eqnarray}
V_{\left( s\right),i^{}_1,i^{}_2,i^{}_3,i^{}_4}^{n^{}_1,n^{}_2,n^{}_3,n^{}_4}=\frac{e^2}{\epsilon^{}_{bp}\ell}
\overline{\sum_{\mathbf{q}}}\frac{V_{11}^s+V_{12}^s}{2N^{}_s}\sum_{j^{}_1,\ldots,j^{}_4}
V_{i^{}_1,i^{}_2,i^{}_3,i^{}_4}^{j^{}_1,j^{}_2,j^{}_3,j^{}_4}\left( \mathbf{q}
\right)&&  \notag \\
\left( u_{n^{}_1,j^{}_1}^{\ast}u^{}_{n^{}_4,j^{}_4}+v_{n^{}_1,j^{}_1}^{\ast}v^{}_{n^{}_4,j^{}_4}\right) 
\left( u_{n^{}_2,j^{}_2}^{\ast }u_{n^{}_3,j^{}_3}+v_{n^{}_2,j^{}_2}^{\ast }v^{}_{n^{}_3,j^{}_3}\right) . &&  \notag
\end{eqnarray}
Note that the inter-layer distance $d$ may not be negligible \cite{luo4},
although $d$ is as small as $2.13$ \AA.

Black phosphorene is also a large $\kappa$ system ($\kappa > 5.6$ when $B<10$
T) due to its heavy effective mass. The mobility of this system is sufficiently high so that 
the $-4/3$ FQHE has already been observed \cite{bpfqheexp}. We check the stability of this odd denominator FQHE for verification of our present approach. With
screening, the $\nu=\pm1/3, \pm4/3$ FQHEs are stable \cite{sm}, which is not surprising and is compatible with the experiment results.
We then explore whether the even denominator FQHE is observable in this material.
Interestingly, the even denominator states $\pm 5/2$ and $\pm 7/2$ FQHEs are not
stable, since the eigen wave function is composed of
different $\psi_{m,i}$. The weights of $\psi_{m\neq 1,i}$ play important roles in
destabilizing these FQHEs.

\begin{figure}[ptb]
\centering
\includegraphics[scale=0.65]{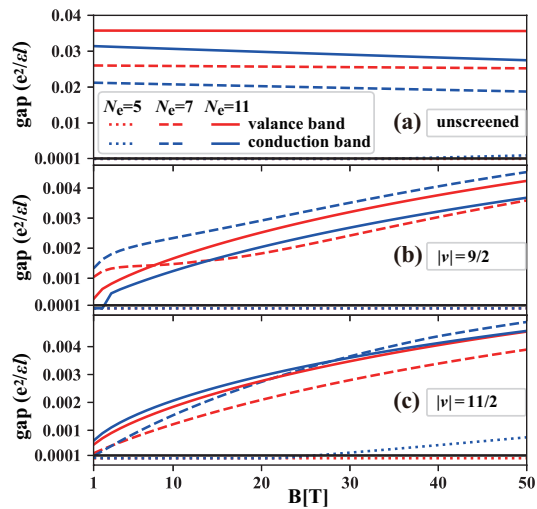}
\caption{ The lowest gaps in the collective modes of the LL $N=2$ black phosphorene states in 
different magnetic fields, (a) without screening, (b) $|\nu|=9/2$ with screening, and (c) 
$|\nu|=11/2$ with screening.
}
\label{fig3}
\end{figure}

For the higher LL $N=2$ ($|\nu|=9/2, 11/2$), the FQHEs survive due to the LL screening. As shown in 
Fig.~\ref{fig3}, the 2DEG may not be incompressible until $B>40$ T if the screening is not considered, 
since the FQHE is not stable for $N^{}_e=5$. However, at $\nu=11/2$ the FQHE can be stabilized for 
a much lower magnetic field ($B>26$ T in the conduction band) for $N^{}_e=5.$ Besides, the gap does 
not change too much when $N^{}_e \ge 7$. For $\nu=9/2$, although the gap decreases a bit when $N^{}_e$ 
is increased from 7 to 11, this gap is still of the order of $10^{-3} {e^2}/{\epsilon^{}_{bp}\ell}$ 
which is one order larger than that in ZnO. We thus expect that $11/2$ FQHE can be
observed, but the mobility of this material may need to be further increased. We also assume that 
the $9/2$ FQHE state is observable (provided the smaller system of $N^{}_e=5$ can be ignored). That
prediction requires experimental verification.


\textit{Summary and remarks}: From the phase diagrams of the ground state at $\nu=5/2$ in the strong LL
mixing region, we find that the two stable FQHE regions (with large energy gap from the
incompressible ground state) for both the parabolic and infinite square well potentials roughly coincide: 
the arc-like region surrounded by low (or negative) energy gap regions. 
This means that the phase transitions between the incompressible phase and the compressible phase occur more than once when either $\kappa$ or width is varied. Moreover, even in the incompressible phase region, the topological phase transition between the (anti-)pfaffian state and the particle-hole symmetric state which is suggested to be stabilized by the LL mixing \cite{feldman}, could be also possible by tuning the magnetic field \cite{luo3} or the width of the quantum well.
The topological features of the ground state should be determined in a thermal 
transport experiment \cite{thermal}. We believe that the particle-hole symmetry \cite{Rezayi}
can be conserved by the extremely strong LL mixing, albeit the topological order of the $5/2$ FQHE 
is still puzzling (especially when $\kappa \sim 1$), since the numerical evidence
has confirmed that the $5/2$ FQHE can be stabilized at shift $-1$ on a
sphere when $\kappa$ is large enough \cite{luo3}. This argument is also compatible
with the screening theory employed here, in which the two-body interaction with renormalized Coulomb 
interaction does not break the particle-hole symmetry.

By combining the screening and finite width corrections, we have successfully explained the latest 
experiments in ZnO. In the strong LL mixing limit, this strategy effectively takes the related 
correlations of other LLs into account and obeys the conservation laws. Moreover, we predict the 
stability of the $5/2$ FQHE and present the phase diagrams at this filling factor for different 
magnetic fields and quantum well widths. The phase diagram should be amenable to verification by 
the experiments.  Another kind of strong LL mixed system, the black phosphorene, is also considered 
and we have shown that the even denominator FQHEs are stabilized in higher LLs by the screening. 
This is also expected to be observed experimentally.
The screening theory is expected to be applicable for study of the FQHE in black arsenic (in the
same nitrogen family as phosphorus) with spin-orbit coupling \cite{blackarsenic} which
also has strong LL mixing, and is probably helpful to understand the even denominator
FQHE in monolayer WSe$_2$ \cite{WSe2}. Indeed, our present approach has laid the foundation for future 
studies of strongly LL mixed systems.


This work has been supported by the NSF-China under Grant No. 11804396. We
are grateful to the High Performance Computing Center of Central South
University for partial support of this work. W.L. thanks Xinyu He for
programming assistance.

\end{document}